\documentclass[aps,twocolumn,pra,showpacs,floatfix]{revtex4}
\usepackage{graphicx}
\usepackage{dcolumn}
\usepackage{amsmath}

\begin{document}

\title{Theoretical study of the experimentally important states of dysprosium}
\author{V. A. Dzuba and V. V. Flambaum}
\affiliation{School of Physics, University of New South Wales,
Sydney 2052, Australia}
\date{\today}

\begin{abstract}

Configuration interaction method is used to calculate transition
amplitudes and other properties of the low states of dysprosium which
are used in cooling and in study of the time variation of the fine
structure constant and violation of fundamental symmetries. 
The branching ratio for the cooling state to decay to states other than
ground state is found to be smaller than $10^{-4}$. The matrix element
of the weak interaction between degenerate states at $E=19797.96$
cm$^{-1}$ is about 2~Hz which is consistent with the experimental
limit $|H_W| = |2.3 \pm 2.9({\rm statistical}) \pm 0.7({\rm
systematic})|$ Hz [A. T. Nguyen, D. Budker, D. DeMille, and M. Zolotorev, 
Phys. Rev. A {\bf 56}, 3453 (1997)] and points to feasibility of its
experimental measurement. Applications include search for physics
beyond the standard model using the PNC isotopic chain approach.
 
\end{abstract}


\maketitle

\section{Introduction}

Atomic dysprosium is proved to be a valuable object to study
fundamental problems of modern physics. It has been recently used in search
for the time variation of the fine structure constant~\cite{Budker07},
study of possible variation of the fine structure constant due to
variation of the gravitation field~\cite{Budker07a}, measurement of
the parity non-conservation (PNC)~\cite{Budker97}, etc. 
Recent progress in trapping and cooling of dysprosium
atoms~\cite{Budker08,Lev,Leefer} opens new exciting possibilities. 

The most interesting feature of dysprosium which has inspired its use in
the study of PNC and variation of the fine structure constant is the
existence of two almost degenerate states of the same 
total momentum and opposite parity. Energy interval is so small that
its actual value is determined by hyperfine structure and isotope
shift. This leads to strong enhancement of both effects.
The states are at the energy $E=19797.96$ cm$^{-1}$ and
both have total momentum $J=10$. Following Ref.~\cite{Budker94} we
use notation A for the even state and notation B for the odd
state. 

Another state of great interest is the odd state at $E=23736.60$
cm$^{-1}$ which is used in cooling. We use notation C for this
state. Our present work mostly focuses on these three states while
some other states are also considered.

The use of the rare-earth atoms, including dysprosium, in atomic PNC
study was first suggested in Ref.~\cite{Dzuba86}, and for the search of the
time-variation of the fundamental constants in
Refs.~\cite{Dzuba99a,Dzuba99b}. 
Dysprosium was studied theoretically in our previous
works~\cite{Dzuba94,Dzuba03,Dzuba08}. The work of 
Ref.~\cite{Dzuba03,Dzuba08} links the change of frequency of the
transition between states A and B to the time-variation of the fine
structure constant. It was used in Ref.~\cite{Budker07,Budker07a} for
the interpretation of the measurements. In Ref.~\cite{Dzuba94} we
calculated the matrix element of the spin-independent parity-violating
weak interaction between states A and B. 

Dysprosium has many stable isotopes and is a good candidate to
study PNC ratio for isotope chains. Such study may reveal important
information on the physics beyond the standard model~\cite{isotopes}.
Accurate atomic calculations of the PNC effect are not needed for this
study. However, reliable estimation of the effect is important to
determine the feasibility of the measurements. 
The result of our previous calculations is
$\langle {\rm A}|W|{\rm B}\rangle$ =  70(40)~Hz~\cite{Dzuba94}. Later
measurements~\cite{Budker97} lead to the limit
$\langle {\rm A}|W|{\rm B}\rangle = |2.3 \pm 2.9({\rm statistical}) \pm
0.7({\rm systematic})$. Although the measured value is not in strong
disagreement with theoretical prediction, given the large uncertainty
of the latter, the experiment did not confirm the large PNC effect
that was hoped for in this system.

In present paper we revisited the PNC calculations and found that
inclusion of more configurations pull 
the value of the weak matrix element down to about 2~Hz which is
consistent with the measurements. The small value of the matrix
element is the result of strong cancellation of different
contributions. Dominant contributions are larger than the final result
by more than order of magnitude. This means that further
cancellation to even smaller number is highly unlikely and the
measurements might be possible on about the same level of sensitivity
which has been already achieved in Ref.~\cite{Budker97}.  
 
Another motivation for this work is due to 
dysprosium cooling at Berkeley~\cite{Budker08,Leefer} and Urbana~\cite{Lev}. 
We study the cooling state C to find the branching ratio
of the transitions from this state to the states other than the
ground state and to the ground state. High value ($\gg 10^{-4}$) for
this ratio would be a problem for cooling. Our present calculations
show that the branching ratio is in fact smaller than $10^{-4}$.

\section{Method}

\begin{table} 
\caption{Configurations and effective core polarizabilities
  $\alpha_p$ (a.u.) used in the calculations.}
\label{t:a}
\begin{ruledtabular}
\begin{tabular}{rlll}
N & \multicolumn{1}{c}{Parity} & 
\multicolumn{1}{c}{Configuration} & 
\multicolumn{1}{c}{$\alpha_p$} \\
\hline
 1 & Even & $4f^{10} 6s^2$   &  0.4 \\     
 2 & Even & $4f^{10} 6s 5d$  &  0.4006 \\     
 3 & Even & $4f^9 6s^2 6p$   &  0.4039 \\
 4 & Even & $4f^9 5d 6s 6p$  &  0.389 \\
 5 & Even & $4f^{10} 6p^2$   &   0.4 \\
 6 & Even & $4f^9 5d^2 6p$   &   0.4 \\
 7 & Odd  & $4f^9 5d^2 6s$   &  0.3947 \\
 8 & Odd  & $4f^9 5d 6s^2$   &  0.3994 \\
 9 & Odd  & $4f^{10} 6s 6p$  &  0.397 \\  
10 & Odd  & $4f^{10} 5d 6p$  &  0.4 \\  
11 & Odd  & $4f^9 5d 6p^2$  &  0.4 \\  
12 & Odd  & $4f^9 6s 6p^2$  &  0.4 \\  
\end{tabular}
\end{ruledtabular}
\end{table}

In present work we use the version of the configuration interaction
(CI) method which was first developed for iron atom~\cite{Dzuba08a}
and then used for other many-electron atoms including
dysprosium~\cite{Dzuba08}. See these works for the detailed
discussion.

The effective Hamiltonian for $N_v$ valence electrons ($N_v$=12 for
dysprosium) has the form
\begin{equation}
  \hat H^{\rm eff} = \sum_{i=1}^{N_v} \hat h_{1i} + 
  \sum_{i < j}^{N_v} e^2/r_{ij},
\label{heff}
\end{equation}
$\hat h_1(r_i)$ is the one-electron part of the Hamiltonian
\begin{equation}
  \hat h_1 = c \mathbf{\alpha \cdot p} + (\beta -1)mc^2 - \frac{Ze^2}{r} 
 + V_{core} + \delta V.
\label{h1}
\end{equation}
Here $\mathbf{\alpha}$ and $\beta$ are Dirac matrixes, $V_{core}$ is
Hartree-Fock potential due to core electrons
and $\delta V$
is the term which simulates the effect of the correlations between core
and valence electrons. It is often called {\em polarization potential} and
has the form
\begin{equation}
  \delta V = - \frac{\alpha_p}{2(r^4+a^4)}.
\label{dV}
\end{equation}
Here $\alpha_p$ is polarization of the core and $a$ is a cut-off parameter
(we use $a = a_B$).

Table~\ref{t:a} lists configurations considered in present work. 
The self-consistent Hartree-Fock procedure is done for every configuration
separately. Then valence states found in the 
Hartree-Fock calculations are used as basis states for the CI calculations.
It is important for the CI method that the atomic core remains
the same for all configurations. We use the core which corresponds to the
ground state configuration. Change in the core due to change of the valence
state is small and can be neglected. This is because core states are not
sensitive to the potential from the electrons which are on large distances
(like $6s$, $6p$ and $5d$ electrons). The $4f$ electrons are on
smaller distances 
and have larger effect on atomic core. However, in all the cases 
(see Table~\ref{t:a}) only one among about ten $4f$ electrons change its state. 
Therefore their effect on atomic core is also small. More detailed
discussion on the effect of valence electrons on atomic core can be 
found in Refs.~\cite{VN,VN1}.

The form of the $\delta V$ in (\ref{dV}) is chosen to coincide with the
standard polarization potential on large distances
($-\alpha_p/2r^4$). We treat $\alpha_p$ as fitting parameters. The
values of $\alpha_p$ 
for each configuration of interest are presented in Table~\ref{t:a}.
They are chosen to fit the experimental position of the configurations
relative to each other. For all configurations the values of
$\alpha_p$ are very close. This is not a surprise since the core is 
the same for every configuration.
Small difference in $\alpha_p$
for different configurations simulates the effect of incompleteness of the
basis and other imperfections in the calculations.

To calculate electric dipole transition amplitudes, lifetimes, magnetic dipole
and electric quadrupole hyperfine structure constants we use the
time-dependent Hartree-Fock method (equivalent of the random phase
approximation) combined with the CI technique, see Ref.~\cite{Ginges}
for a detailed discussion.

\section{Results and discussion}

\begin{table*}
\caption{Experimental and theoretical energies, $g$-factors, hyperfine
  structure constants and lifetimes for some low-lying states of
  $^{163}$Dy. For theoretical uncertainties see discussion in the text.}
\label{t:egab}
\begin{ruledtabular}
  \begin{tabular}{l l c r r r l l r r r r l l}
N &  \multicolumn{1}{c}{leading} &
\multicolumn{1}{c}{Term} &    \multicolumn{1}{c}{$J$} &
\multicolumn{2}{c}{Energies (cm$^{-1}$)} &  
\multicolumn{2}{c}{$g$-factors} &  
\multicolumn{2}{c}{$A$ (MHz)} &  
\multicolumn{2}{c}{$B$ (MHz)} &
\multicolumn{2}{c}{Lifetime}  \\
  &  \multicolumn{1}{c}{config.} & & &
\multicolumn{1}{c}{Expt.\footnotemark[1]} & \multicolumn{1}{c}{Calc.} &
\multicolumn{1}{c}{Expt.\footnotemark[1]} & \multicolumn{1}{c}{Calc.} &
\multicolumn{1}{c}{Expt.} & \multicolumn{1}{c}{Calc.} &
\multicolumn{1}{c}{Expt.} & \multicolumn{1}{c}{Calc.} &
\multicolumn{1}{c}{Expt.} & \multicolumn{1}{c}{Calc.} \\
\hline
\multicolumn{14}{c}{States of special interest}  \\
GS\footnotemark[2] & $4f^{10}6s^2$ & $^5$I &     8 &     0.00 & 0 & 1.24 & 1.24 & 
163\footnotemark[3] & 160 & 1153\footnotemark[3] &  1193 & & \\
A  & $4f^{10}5d6s$ & $^3[10]$ & 10 & 19798 & 19786 & 1.21 & 1.21 &
159\footnotemark[4] & 140 & 1865\footnotemark[4] &  1629 &
7.9~$\mu$s\footnotemark[4]  & 16~$\mu$s \\

B  & $4f^95d^26s$ & $^7$H$^o$ & 10 & 19798 & 19770 & 1.367 & 1.368 &
218\footnotemark[4] & 202 & 2060\footnotemark[4] &  2413 &
$>200~\mu$s\footnotemark[4]  &  0.14 s \\

C  & $4f^{10}6s6p$ & $(8,1)^o$ & 9 & 23737 & 25200 & 1.22 & 1.22 & 
122\footnotemark[5] & 136 & 1842\footnotemark[5] & 2096 &
4.8 ns\footnotemark[4] & 4.7 ns \\

\multicolumn{14}{c}{Some other states}  \\

D & $4f^95d6s^2$ & $^7$H$^o$ & 8 & 7566 & 7563 & 1.35 &1.35 & & 131 &
& 1901 & & 6.9 ms \\ 
E & $4f^95d6s^2$ & $^7$I$^o$ & 9 & 9991 & 9944 & 1.32 & 1.32 & & 125 &
& 2901 & & 3 ms \\
F & $4f^95d6s^2$ & $^5$K$^o$ & 9 & 13496 & 14634 & 1.23 & 1.23 & & 144
& & 4000 & & 21~$\mu$s \\ 
G & $4f^{10}6s6p$ & $(8,2)^o$ & 9 & 17727 & 18092 & 1.25 & 1.26 & & 194
& & 1582 & 2~$\mu$s\footnotemark[4] & 2.9~$\mu$s \\ 

\end{tabular}
\footnotetext[1]{Reference~\cite{Martin}}
\footnotetext[2]{Ground state}
\footnotetext[3]{Reference~\cite{Murakawa,Eliel,Childs}}
\footnotetext[4]{Reference~\cite{Budker94}}
\footnotetext[5]{Reference~\cite{Budker09}}
\end{ruledtabular}
\end{table*}

Table \ref{t:egab} presents the results of calculations for the
energies, $g$-factors, magnetic dipole and electric quadrupole
hyperfine structure constants and lifetimes of some low states of
dysprosium. This includes the degenerate states A and B, cooling state
C and other odd states D to G which may also present an interest for
cooling and quantum information processing~\cite{Lev1}. Calculated values
are compared with available experimental data. Good agreement for the
energies is mostly due to the fitting. Theoretical uncertainty for the
hyperfine structure constants is on the level of 20 to 30\%. The
uncertainty for lifetimes is determined by uncertainties for electric
dipole transition amplitudes. The uncertainties for the amplitudes is
also on the level of 20 to 30\% with the exception of the extremely
small amplitudes ($\ll 1$ a.u.) where uncertainty might be
higher. The amplitudes are presented in Table~\ref{t:dc}.
Maximum disagreement between theory and experiment in
Table~\ref{t:egab} is for the lifetime
of the state A where calculated value 
is two times larger than the experimental one. The lifetimes of the
most of the states are sensitive to the mixing of the $4f^{10}6s6p$
with other odd configurations. For example, the experimental lifetime
of the state C is reproduced in the calculations when state C is
a pure state of the $4f^{10}6s6p$ configuration with very little
admixture of other configurations. This has been achieved by properly
choosing the fitting parameters $\alpha_p$ (see Table~\ref{t:a}). 

\begin{table*}
\caption{Decay channels for states A - G from Table~\ref{t:egab}.}
\label{t:dc}
\begin{ruledtabular}
  \begin{tabular}{c l c r r r c c} 
\multicolumn{1}{c}{Upper} & \multicolumn{3}{c}{Lower state} &
\multicolumn{1}{c}{Energy} &
\multicolumn{1}{c}{$\omega$} &
\multicolumn{1}{c}{$|\langle f||\mathbf{D}||i \rangle|$} &
\multicolumn{1}{c}{Probability} \\
\multicolumn{1}{c}{state} & \multicolumn{1}{c}{Config.} &
\multicolumn{1}{c}{Term} & \multicolumn{1}{c}{$J$} &
 \multicolumn{1}{c}{cm$^{-1}$} &
 \multicolumn{1}{c}{cm$^{-1}$} &
\multicolumn{1}{c}{(a.u.)} & \multicolumn{1}{c}{s$^{-1}$} \\
\hline
A & $4f^95d6s^2$  & $^7$K$^o$ & 10 & 12893 &  6905 & 0.056 & 0.982E+02\\
  & $4f^{10}6s6p$ & $(8,2)^o$ & 10 & 17513 &  2285 & 0.276 & 0.874E+02\\
  & $4f^95d6s^2$  & $^7$I$^o$ & 9 &   9991 &  9807 & 0.049 & 0.218E+03\\
  & $4f^95d6s^2$  & $^5$K$^o$ & 9 &  13496 &  6302 & 0.192 & 0.887E+03\\
  & $4f^{10}6s6p$ & $(8,1)^o$ & 9 &  15972 &  3826 & 2.34 & 0.296E+05\\
  & $4f^95d6s^2$  & $^7$K$^o$ & 9 &  16717 &  3080 & 0.441 & 0.547E+03\\
  & $4f^{10}6s6p$ & $(8,2)^o$ & 9 &  17727 &  2071 & 5.86 & 0.294E+05\\

B & $4f^{10}5d6s$ & $^3[9]$   & 10 & 18463 & 1335 & 0.069 & 0.109E+01\\
  & $4f^{10}5d6s$ & $^3[8]$   &  9 & 17515 & 2283 & 0.073 & 0.608E+01\\
  & $4f^{10}5d6s$ & $^3[9]$   &  9 & 19241 &  557 & 0.0045 & 0.339E-03\\

C & $4f^{10}6s^2$ & $^5$I     &  8 &     0 & 23737 & 12.28  &  0.215E+09 \\
  & $4f^{10}5d6s$ & $^3[7]$   &  8 & 17613 &  6123 &  0.221 &  0.120E+04 \\
  & $4f^{10}5d6s$ & $^3[8]$   &  8 & 18903 &  4833 &  0.265 &  0.845E+03 \\
  & $4f^{10}6s^2$ & $^3$K2    &  8 & 19019 &  4717 &  0.057 &  0.362E+02 \\
  & $4f^{10}5d6s$ & $^3[9]$   &  8 & 20194 &  3543 &  0.036 &  0.601E+01 \\
  & $4f^96s6p$ & $(^{15}/_2,^1/_2)$  &  8 & 20790 &  2947 &  0.099 &  0.265E+02 \\
  & $4f^{10}5d6s$ & $^3[7]$   &  8 & 21603 &  2134 &  0.226 &  0.530E+02 \\
  & $4f^{10}5d6s$ & $^3[8]$   &  9 & 17515 &  6222 &  0.411 &  0.434E+04 \\
  & $4f^{10}5d6s$ & $^3[9]$   &  9 & 19241 &  4496 &  0.122 &  0.144E+03 \\
  & $4f^{10}5d6s$ & $^3[10]$  &  9 & 20209 &  3528 &  0.222 &  0.231E+03 \\
  & $4f^{10}5d6s$ & $^1[9]$   &  9 & 22046 &  1691 &  0.659 &  0.224E+03 \\
  & $4f^{10}5d6s$ &           &  9 & 23218 &   518 &  0.0076 &  0.846E+03 \\
  & $4f^{10}5d6s$ & $^3[9]$   & 10 & 18462 &  5274 &  0.502 &  0.394E+04 \\
  & $4f^{10}5d6s$ & $^3[10]$  & 10 & 19798 &  3939 &  0.467 &  0.142E+04 \\
  & $4f^{10}5d6s$ & $^1[10]$  & 10 & 22487 &  1249 &  1.039 &  0.225E+03 \\

D & $4f^{10}6s^2$ & $^5$I     &  8 &     0 &  7566 &  0.053 &  0.146E+03 \\
  & $4f^{10}6s^2$ & $^5$I     &  7 &  4134 &  3431 &  0.0017 &  0.136E-01 \\

E & $4f^{10}6s^2$ & $^5$I     &  8 &     0 &  9991 &  0.059 &  0.369E+03 \\
F & $4f^{10}6s^2$ & $^5$I     &  8 &     0 & 13496 &  0.424 &  0.471E+05 \\
G & $4f^{10}6s^2$ & $^5$I     &  8 &     0 & 17727 &  0.897 &  0.478E+06 \\
\end{tabular}
\end{ruledtabular}
\end{table*}

Table~\ref{t:dc} shows decay channels for the states listed in
Table~\ref{t:egab}. The data is based on the calculated electric
dipole transition amplitudes and experimental energies. No electric
quadrupole or magnetic dipole amplitudes were taken into account. Test
calculations show that their contribution is negligible. The analysis
of the data for state A shows that it decays mostly to two states of
the $4f^{10}6s6p$ configuration. This is because the leading
configuration of state A is the $4f^{10}6s5d$ configuration and an
electric dipole transition between the two states can be reduced to
the allowed single-electron $6p \rightarrow 5d$ transition. In
contrast, transitions to the states of the $4f^95d6s^2$ configuration
can only go via configuration mixing. The calculated lifetime of
the state A is larger than the experiment. This means that the
transition amplitudes are smaller. It is unlikely that the amplitudes
of the transitions between state A and states of the $4f^{10}6s6p$
configuration are underestimated. This is because the latter states
are almost pure, with only about 3\% admixture of other
configurations. It is more likely that some of the other amplitudes
are underestimated because of too small mixture of the $4f^{5}5d6s^2$
and $4f^{10}6s6p$ configurations. 

The analysis of the decay channels of the cooling level C (see
Table~\ref{t:dc}) also indicates
the sensitivity of the data to the mixing of the $4f^{5}5d6s^2$ and
$4f^{10}6s6p$ configurations. Good agreement with experiment for the
lifetime of the state C can only be achieved if this state is a pure
state of the $4f^{10}6s6p$ configuration. On the other hand, this
state is very close to the states of the $4f^{5}5d6s^2$
configuration. Manipulating with the fitting parameters $\alpha_p$
(Table~\ref{t:a}) can easily lead to a situation when the states of
the two configurations are strongly mixed and the lifetime of the 
state C is larger than the experiment. Therefore, the lifetime of this
state should be monitored in the fitting process. 

The data in Table~\ref{t:dc} allows to estimate the ratio of the
following probabilities: (a) decay of state C to all lower state other
than the ground state, and (b) decay of the state C into the ground
state. This ratio is $(6 \pm 3) \times 10^{-5}$. 
The fraction of atoms lost into metastable states must be
even smaller because the number above includes all channels of
decay without considering which of them end up in a metastable state.
However, some lower states quickly decay to the ground state and do
not produce any losses.
A more detailed analysis would require a lot of extra work, however it
is not needed since the branching ratio
$<10^{-4}$ is sufficiently good for the cooling~\cite{Lev}. 

Table~\ref{t:dc} also shows the data for states D-G which might be
useful for cooling and information processing~\cite{Lev1}.

Our calculated value for the electric dipole reduced matric element
between states A and B is $0.024\pm 0.010$ a.u. This agrees well with
the experimental value of 0.015(1) a.u. from Ref.~\cite{Budker94}.
Note that this amplitude is zero in the non-relativistic limit. This
is because states A and B have different spin (see Table~\ref{t:egab})
and electric dipole operator cannot change it. In relativistic
calculations the amplitude is not zero. However, it is small and this
small value is the result of strong cancellations between different
contributions. Table~\ref{t:e1} shows the largest in absolute values
contributions to the electric dipole transition amplitude between
states A and B. All these contributions are due to the $4f$ - $5d$
electric dipole transitions within the main configurations of states A
and B. The sum is only 0.0017 a.u. which is more than an order of
magnitude smaller than the final answer. The final amplitude is the
sum of many smaller contributions which contain all possible
single-electron transitions.

\begin{table}
\caption{Contributions to the electric dipole transition amplitude
  between states A and B (a.u.)}
\label{t:e1}
\begin{ruledtabular}
  \begin{tabular}{l l c r r}

\multicolumn{2}{c}{Configurations} & 
\multicolumn{1}{c}{Single-electron} & 
\multicolumn{1}{c}{Partial} & 
\multicolumn{1}{c}{Sum} \\ 
\multicolumn{1}{c}{State A} & 
\multicolumn{1}{c}{State B} & 
\multicolumn{1}{c}{matrix element} & 
\multicolumn{1}{c}{contribution} & 
\multicolumn{1}{c}{(a.u.)} \\ 
\hline
$4f^{10}5d6s$ & $4f^95d^26s$ & $\langle 4f_{5/2}|D| 5d_{3/2}\rangle$ & -0.2634 & -0.2634 \\
$4f^{10}5d6s$ & $4f^95d^26s$ & $\langle 4f_{5/2}|D| 5d_{5/2}\rangle$ &-0.0106 & -0.2740 \\
$4f^{10}5d6s$ & $4f^95d^26s$ & $\langle 4f_{7/2}|D| 5d_{5/2}\rangle$ & 0.2723 & -0.0017 \\
\end{tabular}
\end{ruledtabular}
\end{table}

The situation is even more complicated for the matrix element of the
weak interaction between states A and B. Table~\ref{t:pnc} shows
dominant contributions to this matrix element. There are strong
cancellations between different contributions, mostly between terms
containing the $\langle 4f_{5/2}|H_W| 5d_{5/2}\rangle$ and
$\langle 6s_{1/2}|H_W| 6p_{1/2}\rangle$ single-electron matrix
elements. The former of these matrix elements is small. It is not zero due
to the effect of core polarization by the weak interaction: weak
interaction acts on $s$ and $p$ core electrons changing the
self-consistent Hartree-Fock potential which in turn leads to the $4f$
- $5d$ transition between valence states. The $\langle 6s_{1/2}|H_W|
6p_{1/2}\rangle$ integral is not small but its contribution is
suppressed by small admixture of the configurations containing the $6s$
and $6p$ states to the main configurations. This makes the result 
very sensitive to configuration mixing. We found that inclusion
of all configurations listed in Table~\ref{t:a} is important for the
weak matrix element. The effect of some configurations is indirect,
via changing the coefficients of configuration mixing. For example,
the sum of all largest contributions listed in Table~\ref{t:pnc},
-2.44~Hz is very close to the final answer -2.13~Hz. This may make an
impression that the configurations not listed in Table~\ref{t:pnc} do
not contribute. However, if e.g. the $4f^96s^26p$ configuration is not
included, the weak matrix element becomes larger by more than four
times. This configuration was missed in our previous
calculations~\cite{Dzuba94} which is probably one of the reasons of
disagreement between theory and experiment. The result of
Ref.~\cite{Dzuba94} corresponds to the situation when the weak matrix 
element between states A and B is dominated by the contribution of the  
$\langle 4f_{5/2}|H_W| 5d_{5/2}\rangle$ single-electron matrix element
(first line of Table~\ref{t:pnc}). This is in spite of the fact that
most of the other configurations were also included. Probably
incomplete inclusion of other configurations leads to underestimation
of the configuration mixing which suppresses the
contributions containing other single-electron matrix elements.

Our present result for the weak matrix element (2~Hz)
is consistent with the experimental value of  $|\langle {\rm A}|H_W|{\rm
B}\rangle| = |2.3 \pm 2.9({\rm statistical}) \pm 0.7({\rm
systematic})|$~\cite{Budker97}. This is about 20 times
smaller than the values of individual contributions (see
Table~\ref{t:pnc}). Since there is no law of physics to make this
matrix element exactly zero we believe that any further
cancellation which would make the result even smaller than 2~Hz is
highly unlikely. This probably means that the measurements of the
PNC effect can be done on about the same level of sensitivity which
has been already achieved in Ref.~\cite{Budker97}.

\begin{table}
\caption{Contributions to the matrix element of the weak interaction
  between states A and B (Hz)}
\label{t:pnc}
\begin{ruledtabular}
  \begin{tabular}{l l c r r}

\multicolumn{2}{c}{Configurations} & 
\multicolumn{1}{c}{Single-electron} & 
\multicolumn{1}{c}{Partial} & 
\multicolumn{1}{c}{Sum} \\ 
\multicolumn{1}{c}{Left} & 
\multicolumn{1}{c}{Right} & 
\multicolumn{1}{c}{matrix element} & 
\multicolumn{1}{c}{contribution} & 
\multicolumn{1}{c}{(Hz)} \\ 
\hline
$4f^{10}5d6s$ & $4f^95d^26s$ & $\langle 4f_{5/2}|H_W| 5d_{5/2}\rangle$ & 38.79 &  38.79 \\
$4f^{10}5d6s$ & $4f^{10}5d6p$ & $\langle 6s_{1/2}|H_W| 6p_{1/2}\rangle$ &-49.04 & -10.25 \\
$4f^{10}5d6s$ & $4f^{10}6s6p$ & $\langle 5d_{3/2}|H_W| 6p_{3/2}\rangle$ &  1.60 & -8.65 \\
$4f^95d6s6p$  & $4f^95d6s^2$ & $\langle 6p_{1/2}|H_W| 6s_{1/2}\rangle$ & -1.35 & -10.00 \\
$4f^95d6s6p$  & $4f^95d^26s$ & $\langle 6p_{3/2}|H_W| 5d_{3/2}\rangle$ & -4.28 & -14.27 \\
$4f^95d^26p$  & $4f^95d^26s$ & $\langle 6p_{1/2}|H_W| 6s_{1/2}\rangle$ & 11.84 &  -2.44 \\
\end{tabular}
\end{ruledtabular}
\end{table}


\section*{Acknowledgments}

The authors are grateful to Nathan Leefer, Dmitry Budker and Benjamin
Lev for stimulating discussions.
The work was funded in part by the Australian Research Council.

\end{document}